\documentclass[%
reprint,
 amsmath,amssymb,
 prb,
]{revtex4-1}
\pdfoutput=1
\newcommand{\be}{\begin{eqnarray}}
\newcommand{\ee}{\end{eqnarray}}

\usepackage[pdftex]{graphicx}

\newcommand{\bra}[1]{\left \langle{#1}\right |}
\newcommand{\ket}[1]{\left |{#1}\right \rangle}
\usepackage{dcolumn}
\usepackage{bm}
\usepackage{multirow}
\newcommand{\rom}[1]{\uppercase\expandafter{\romannumeral #1\relax}}

\begin{document}

\preprint{APS/123-QED}
\title{Universal logic gates for quantum-dot-electron-spin qubits using trapped quantum-well exciton-polaritons}

\author{Shruti Puri$^1$, Peter L. McMahon$^1$ and Yoshihisa Yamamoto$^{1,2}$ }
\affiliation{1. E.\,L. Ginzton Laboratory, Stanford University, Stanford, California 94305, USA}
\affiliation{2. National Institute of Informatics, 2-1-2 Hitotsubashi, Chiyoda-ku, Tokyo 101-8430, Japan}%

\date{\today}
\begin{abstract}
In this paper we introduce and analyze a new system design for quantum-dot-based qubits that simultaneously supports scalable one-qubit and two-qubit gates, and single-shot qubit measurement. All three key processes (one-qubit gates, two-qubit gates, and qubit measurement) rely on the interaction between the electron in each quantum dot and exciton-polaritons formed in a quantum well situated near the quantum dots. A key novel feature of our proposed system is the use of polariton traps, which we show enhances the quantum-dot--quantum-well interaction by a factor of 10 and consequently results in $100 \times$ faster two-qubit gates. We also introduce a novel one-qubit gate that is based on a combination of optical and microwave control, which is supported in the same device and system configuration as the other operations, in contrast to the conventional one-qubit gate that is based on all-optical control.


\end{abstract}

\maketitle

\section{Introduction}


The construction of a large-scale quantum computer requires the identification of scalable qubits with long coherence times and support for universal quantum gates, initialization, and readout (the DiVincenzo criteria\cite{divincenzo2000physical}). Many choices of gate sets are universal, but they share in common the feature that both single-qubit and multi-qubit gates are needed, and the simplest choices typically involve a finite set of single-qubit gates and one two-qubit entangling gate (for example \cite{jones2012faster}, $\left\{\mathtt{X},\mathtt{Y},\mathtt{Z},\mathtt{H},\mathtt{S},\mathtt{T},\mathtt{CNOT}\right\}$). When considering the implementation of a fault-tolerant error correction scheme, such as the surface code \cite{fowler2012surface}, smaller (non-universal) gate sets at the physical layer can be sufficient \cite{jones2012layered}, but still a set of one-qubit gates and an entangling two-qubit gate are required (for example \cite{jones2012layered}, $\left\{\mathtt{X},\mathtt{Y},\mathtt{Z},\mathtt{H},\mathtt{CNOT}\right\}$). The net implication for designing and evaluating physical systems for use as qubits is that it is necessary to produce a design that simultaneously allows both single-qubit gates and an entangling two-qubit gate to be performed. The design should also allow for readout and initialization of each qubit, which can respectively be realized by quantum non-demolition (QND) measurement, and QND measurement followed by a conditional $\mathtt{X}$ operation. The measurement operation should be single-shot\cite{puri2014single,elzerman2004single,vijay2011observation,morello2010single,vamivakas2010observation,sun2016single}.

Any successful candidate platform for large-scale quantum computation using surface code error correction must meet the threshold requirements for fault-tolerant operation, and must be sufficiently scalable. One scenario, worked out in detail, for the construction of a computer capable of factoring a 1024-bit integer calls for the following~\cite{jones2012layered}: (i) a 2D array of $~10^8$ physical qubits (ii) error probability after each gate operation $<1\%$, and (iii) gate time and measurement time $\sim$ 10's of ns. Self-assembled semiconductor quantum dots are one candidate platform that can offer the prospect of scaling to such large numbers of physical qubits, and it is possible to grow large arrays of site-controlled QDs ~\cite{schneider2008lithographic}. Spins in optically-active quantum dots have been extensively studied as potential qubits both theoretically and experimentally over the past two decades. However, it has proven extremely challenging to find a system design in which all the criteria for fault-tolerant, universal operation can be satisfied simultaneously. Single-qubit gates for quantum-dot qubits implemented via picosecond optical pulses are well-established theoretically and experimentally \cite{de2013ultrafast,press2008complete}. Unfortunately most recent theoretical proposals for scalable two-qubit gates and for single-shot QND measurement make assumptions about the system design that are incompatible with the established one-qubit gates, which we will elaborate on shortly.

Furthermore, a scalable two-qubit gate, which requires an interaction between neighboring quantum dot electron spins, is yet to be demonstrated. One approach to constructing a two-qubit gate is based on the dispersive interaction of cavity photons with two electron spins~\cite{imamog1999quantum,ladd2011simple}. The performance of this gate is characterized by the cooperativity factor $C$ of the cavity, which is proportional to the ratio of its quality factor $Q$ and mode volume $V$. It is possible to reach error rates below the fault-tolerance threshold using cavities with large cooperativity factor, $C>10^3$. Although a planar microcavity has a large extent (the size of the chip), the cavity modes have a much smaller transverse extent~\cite{ujihara1994simple,osuge1994spontaneous}. The inherent mode radius in a planar microcavity is: $R=\sqrt{\lambda L_\mathrm{c}/2\pi(1-r_1r_2)}$, where $\lambda$ is the optical wavelength, $L_\mathrm{c}$ is the effective cavity length and $r_1(r_2)$ is the reflectivity of the top(bottom) mirror~\cite{bjork1993spontaneous}. If the mirror reflectivities are increased to increase the quality factor, the mode radius (equivalently the mode volume) also increases. As a result it is difficult to achieve a high cooperativity factor in planar microcavities. In summary, despite the passage of more than 15 years since the first dispersive two-qubit gate proposal~\cite{imamog1999quantum}, not even an unscalable proof-of-principle demonstration (for example, in a micropillar or microdisk cavity) has been performed, and it seems unlikely that a two-qubit gate will be realized with this technique when a planar cavity is used, which is necessary for the system to be scalable.

Fast, high-fidelity, single-shot measurement of the qubit state is also yet to be realized in a scalable single-QD system. One prominent measurement scheme is based on the qubit-spin-dependent Faraday rotation of the cavity output field~\cite{atature2007observation}. This technique suffers from excess measurement backaction: achieving a regime in which the probe pulse is strong enough to yield single-shot operation, but weak enough to avoid causing spin-flip Raman transitions (with probability $\propto 1/C$)~\cite{puri2014single} appears to be difficult.

\section{QW-polariton based approach}

There has been a series of proposals to introduce a quantum well (strongly coupled to a surrounding microcavity) beneath the quantum dots and use the spin-dependent exchange interaction between the quantum-well exciton-polaritons and the quantum dot electrons to realize two-qubit gates and QND measurement \cite{quinteiro2006long,puri2012two,puri2014single}. A convenient feature of this approach is that both two-qubit gates and measurement are supported in the same device and experimental setup. However, these proposals all assume that the magnetic field (used to split the ground spin states in energy) is applied in the Faraday geometry (magnetic field parallel to the sample growth direction) ~\cite{de2013towards}, whereas the conventional one-qubit gate assumes the Voigt geometry (magnetic field perpendicular to the sample growth direction) ~\cite{de2013towards}. Consequently although two-qubit gate operations and measurement are supported concurrently in these coupled quantum-well--quantum-dot devices, one-qubit optical gates in the vein of Ref. \cite{press2008complete} appear to be disallowed so long as Faraday geometry is assumed.

We note that in the two previous proposals of polariton-mediated interaction between spin qubits ~\cite{quinteiro2006long,puri2012two}, the analyses were performed for systems other than self-assembled quantum dots: in the former case, donor spins were explored, and in the latter case, electrostatically-defined quantum dot were the system under study. These settings are indeed similar to the case of self-assembled quantum dots, but several important system parameters are different. In this paper we analyze in detail the case of self-assembled quantum dots, and extend the analysis of Ref. ~\cite{puri2012two} to this system, using realistic device and experimental parameters to obtain estimates of fidelities and gate times.

In this paper we introduce two new aspects to the system design of a qubit platform based on quantum dots with quantum-well polaritons: polariton traps, and a one-qubit gate based on a combination of direct radiofrequency manipulation and polariton-based resonance tuning. In so doing we find we are able to improve the previous polariton-based two-qubit and measurement operations, and also design a platform that simultaneously supports all the operations to satisfy the DiVincenzo criteria.

Throughout this paper we propose using a control mechanism based on the Coulomb exchange interaction between a QD electron spin and optically excited, 2D microcavity exciton-polaritons that are laterally confined in a micron-sized quantum well trap (QWT). Polariton-based schemes for quantum dot qubit control \cite{quinteiro2006long,puri2012two,puri2014single} generally offer two advantages: firstly, it is possible to design the QW and QDs such that the optical fields applied to implement operations are far-detuned from the direct QD optical transitions, thus preventing any unwanted backaction. Secondly, the bosonic nature of polaritons and their weak interaction with the solid-state environment allows the injection of numerous polaritons coherently in a single mode, increasing the nonlinearity in the qubit-polariton coupling, crucial for two-qubit operation. In this paper we find that the lateral confinement of the polaritons in the ``traps'' increases the Coulomb exchange energy and eliminates the dependence between $Q$ and $R$. Unfortunately the extension of the previous two-qubit gate proposals \cite{quinteiro2006long,puri2012two} in this new device design is not entirely straightforward, but we nevertheless are able to present a pulse protocol that can realize a two-qubit gate using trapped polaritons. We also analyze the previous polariton-mediated QND measurement scheme \cite{puri2014single} in this trapped-polariton setting, and introduce a polariton-controlled one-qubit gate.
\\
\section{Setup}
As illustrated in Fig.~1, the setup consists of a 2D array of self-assembled In$_x$Ga$_{1-x}$As QDs grown on top of a In$_y$Ga$_{1-y}$As QW, with a few-monolayer-thick GaAs barrier in-between them. The QW can be grown 4-8 nm thick, which is of the order of an exciton Bohr radius. The QDs are pyramidal or lens-like 3D islands with a typical height in the range $\sim1-4$ nm and base width in the range $\sim20-50$ nm. A single electron is trapped in each of the QD. The QD electron spin is quantized along the growth ($z$) axis by an external magnetic field $B_0\hat{z}$, a configuration which is known as the Faraday geometry. The QDs and QW are embedded in a GaAs $\lambda$ cavity formed by AlGaAs/AlAs distributed Bragg mirrors (DBRs). 

In order to laterally confine the polaritons, the GaAs spacer region is etched before the growth of the top DBR, creating a locally ($\Delta\lambda_\mathrm{c} \sim 5$ nm) thicker cavity~\cite{el2006polariton}. The length of the cavity in the etched region is $\lambda_\mathrm{c}$ and that in the spacer is $\lambda_\mathrm{c}-\Delta\lambda_\mathrm{c}$. This local modulation of cavity length introduces a microscale trap potential $\hbar\omega_\mathrm{t}={2\pi \hbar c\Delta\lambda_\mathrm{c}}/{n\lambda_\mathrm{c}L_\mathrm{c}}\sim7$ meV for cavity photons~\cite{lugan2006theory}. Here $L_\mathrm{c}$ is the cavity thickness including the penetration depth in the DBR. The photons outside the trap region are non-resonant with the cavity and have an extremely short lifetime. In the etched region, the QW exciton is resonant with the lowest cavity photon mode (${\bf{k}}_{||}={\bf{0}}$) and in the strong coupling regime, the resulting eigenmodes of the system are upper polaritons (UPs) and lower polaritons (LPs) \cite{lugan2006theory,el2006polariton,muller2006high,weisbuch1992observation}. The splitting between the UP-LP mode ($2\Omega_\mathrm{R}$) depends on the strength of the dipole coupling between the QW excitons and cavity photons and typically, in a single QW, $2\Omega_\mathrm{R}\sim 3-4$ meV. Furthermore, the lateral confinement of the polaritons, defined by electron-beam lithography \cite{el2006polariton,muller2006high}, results in the discretization of the energy-momentum dispersion so that the LP mode with zero in-plane momentum, ${\bf{k}}_\|={\bf{0}}$ is the ground state. The splitting between the ground state and first excited state of the LP in a trap of radius $R\sim 1$ $\mu$m is $\sim 1$ meV, making it possible to selectively excite only the ground state polaritons with ${\bf{k}}_\|={\bf 0}$. 

In the trap, the polarization of the LPs is quantized along the growth direction. The LPs with angular momentum $J_\mathrm{z}=1$ or -1 comprise of an electron with $s_{\mathrm{ze}}=-\frac{1}{2}$ or $+\frac{1}{2}$ and a heavy-hole with $(l_{\mathrm{zhh}},s_{\mathrm{zhh}})=(1, \frac{1}{2})$ or $(-1,-\frac{1}{2})$, where $s$ and $l$ refer to spin and orbital angular momentum \cite{dyakonov2008spin,shelykh2005spin}. Because of the QW exciton selection rules, left $(\sigma^+)$ or right circularly polarized $(\sigma^-)$ laser pulse selectively excites LPs with $J_\mathrm{z}=+1$ or $-1$ respectively~\cite{deng2003polariton}. Next we describe how these LPs can tunnel between neighbouring traps.

\begin{figure}
\includegraphics[width=8cm,height=5cm]{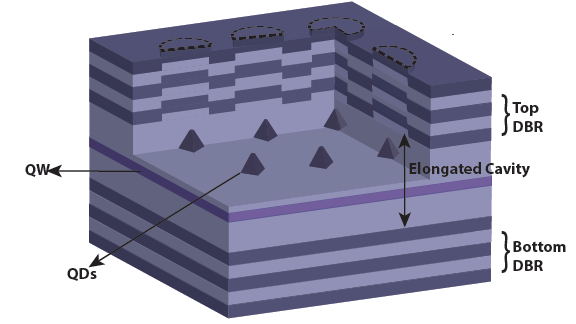}\\
\caption{Illustration of the proposed setup: 2D array of QDs coupled to QW in patterned DBR microcavity. The cavity length is locally modulated to create potential traps for photons. }
\end{figure}

\section{Tunnel Coupling Between Neighboring Traps}
Our aim is to use the optically-excited QW LP mode to control and manipulate the QD electron spin qubits. For controlling a single electron spin trapped in a QD, LPs must be excited in the trap under that QD. However, a two-qubit operation requires two coupled ${\bf{k}}_\|={\bf{0}}$ LP modes in the traps under two adjacent QDs. The Hamiltonian for this linear tunnel coupling between the LP modes in two neighboring traps is given by:
\be
\hat{H}_\mathrm{t.c.}=\hbar \omega_1 a^\dag_1 a_1+\hbar \omega_2 a^\dag_2 a_2+U(a^\dag_1 a_2+a^\dag_2 a_1)\nonumber
\ee
where, $\hbar \omega_{1}$ and $\hbar \omega_{2}$ is the energy of the LP mode in trap 1 and 2 respectively, $a^\dag_{i}$, $a_{i}$ ($i=1,2$) is the creation, annihilation operators for the LP mode in trap $i$ and $U$ is the coupling strength. It is possible to estimate $U$ by solving the time-independent Schr\"{o}dinger's equation for a particle of effective mass of the LP $m_\mathrm{LP}$ in a double potential well of depth given by the trap potential $\hbar\omega_\mathrm{t}$ separated by a distance $D$~\cite{sarchi2008coherent} (Appendix A). Since, the in-plane effective mass of LPs is $\sim 10^3$ smaller than that of excitons~\cite{deng2010exciton}, it is possible to achieve a strong tunnel coupling between the two LP modes even when the distance between them, $D\sim 1-2$ $\mu$m. For example, the coupling strength between two traps of half-width $R=1$ $\mu$m that are separated by $D=0.5$ $\mu$m (or 2.5 $\mu$m center-to-center) is $U=0.5$ meV. A strong coupling between traps is essential for a two-qubit operation, but it can also lead to crosstalk. Next, we examine a protocol to eliminate this crosstalk.\\

Consider a target trap T, which is directly under the QD that hosts the spin qubit we intend to manipulate. In a 2D lattice, this trap has four nearest neighbours to which it can be coupled. An input pump of flux $F_\mathrm{\tiny{T}}$ and $F$ excites the LPs in the target trap and its nearest neighbours respectively. Both the pumps are red detuned from the ${\bf{k}}_\|={\bf{0}}$ LP mode in the traps by $\delta$. There are two possible injection paths in the neighbouring trap, 
\begin{enumerate}
\item Direct excitation by the pump at rate $F$, and  
\item Indirect injection from trap T via tunnel coupling at a rate $=U\times$ amplitude of coherent field inside the target trap=$UF_\mathrm{T}/(\delta-i\gamma/2)$.
\end{enumerate}
As a result if $F=UF_\mathrm{T}/(\delta-i\gamma/2)$, these two paths interfere destructively and the neighbouring traps are left empty. A more rigorous derivation has been provided in Appendix B. In this way it is possible to minimize unwanted LP tunnelling between the traps and eliminate crosstalk.

\section{Control Mechanism}
\subsection{Coulomb Exchange Interaction}
Careful design of the GaAs barrier thickness and In concentrations of the QD and QW ($x$ and $y$), results in a band structure such that the electron is primarily confined in the QD but has a non-zero wavefunction in the QW. The finite overlap of the localized QD electron and 
the LPs in the trap directly below the QD, results in a spin-dependent Coulomb exchange interaction between them \cite{piermarocchi2002optical,quinteiro2006long}. It has been shown that in the low density limit this interaction is represented by the Hamiltonian \cite{puri2012two},
\be 
H_\mathrm{I}&=&V_{\mathrm{ex}} (a^\dag_{-1}a_{-1}- a^\dag_1 a_1){\sigma}_\mathrm{z} 
\label{hintred}
\\
V_{\mathrm{ex}}&=&|r_0|^2\int d{\bf{r}}_\mathrm{e}d{\bf{r}}_\mathrm{2}d{\bf{r}}_\mathrm{1}\frac{\psi({\bf{r}}_\mathrm{e},{\bf{r}}_\mathrm{2})\phi({\bf{r}}_\mathrm{1})e^2\psi({\bf{r}}_\mathrm{1},{\bf{r}}_\mathrm{2})\phi({\bf{r}}_\mathrm{e})}{4\pi\epsilon(|{\bf{r}}_\mathrm{e}-{\bf{r}}_\mathrm{1}|)}, \nonumber\\
\ee
where $a^\dag_{1,-1} (a_{1,-1})$ are the creation (annihilation) operators for the LP mode with ${\bf{k}}_\|={\bf{0}}$ and $J_\mathrm{z}=\pm 1$, $\epsilon$ is the dielectric constant of the In$_y$Ga$_{1-y}$As QW, ${\bf{r}}_\mathrm{1},{\bf{r}}_\mathrm{2}$ are the position vectors of the electron and hole in the excitonic part of the LP, ${\bf{r}}_\mathrm{e}$ is that of the QD electron, $\psi$ and $\phi$ represent the wavefunctions of the excitonic component of the LP and of the localized electron, ${{\sigma}_\mathrm{z}}$ is the Pauli spin operator of the QD electron. $r_0$ is the excitonic Hopfield coefficient of ${\bf{k}}_{||}={\bf{0}}$ LPs~\cite{hopfield1958theory}. When the cavity photons and QW excitons are resonant at ${\bf{k}}_{||}={\bf{0}}$, $r_0=1/\sqrt{2}$. From Eq.~\eqref{hintred} we see that the exchange interaction induces a spin-dependent shift in the LP resonance. If the QD spin state is $\ket{{s}_{\mathrm{z}}}=\ket{1/2}$, then the resonance energy of a $J_\mathrm{z}=-1(+1)$ LP will decrease (increase) by an amount $V_{\mathrm{ex}}$, making the $J_\mathrm{z}=1$ and $J_\mathrm{z}=-1$ LP non-degenerate. This effect is reversed if $\ket{{s}_{\mathrm{z}}}=\ket{-1/2}$. These spin-dependent shifts of the LP resonance can be employed to achieve qubit operations. It has been shown that $V_\mathrm{ex}\propto 1/A$, where $A$ is the area of the region in which the LPs are excited~\cite{puri2014single} and is independent of the polariton lifetime. As a result, we can obtain a large exchange energy by exciting LPs in a small area without decreasing their lifetime~\cite{muller2006high}. For a typical trap of radius $R=1$ $\mu$m, $A=2.2$ $\mu$m$^2$ and if $x=30\%$, $y=15\%$, the size of the QD is 20 nm $\times$ 20 nm $\times$ 1.5 nm, the QW thickness is 6 nm, and the barrier layer is 1 nm thick, then we estimate that $V_{\mathrm{ex}}\approx 2$ $\mu$eV~\cite{puri2014single}. 

\subsection{Dipole Coupling}
In addition to the above Coulomb exchange coupling, the photonic part of the LP couples to the QD single electron-trion transition via dipole coupling. LPs with $J_\mathrm{z}=1$ couple the single electron with spin $s_\mathrm{z}=1/2$ to the trion with total angular momentum $J_\mathrm{z}=3/2$. Similarly, LPs with $J_\mathrm{z}=-1$ couple the single electron with spin $s_\mathrm{z}=-1/2$ to the trion with total angular momentum $J_\mathrm{z}=-3/2$~\cite{press2008complete}. The band structure of the proposed sample is designed such that this coupling is off-resonant resulting in the Stark shift of the QD resonance~\cite{berezovsky2008picosecond},
\be
H_\mathrm{d}=-\chi(a^\dag_1a_1-a^\dag_{-1}a_{-1})\sigma_\mathrm{z},\quad \chi=\frac{t^2_0g^2}{2\Delta}
\ee
where $t_0$ is the fraction of photonic component in the LP, $g$ is the dipole coupling constant and $\Delta$ is the detuning between the LP resonance and the QD single-electron trion transition. In our system we estimate the Stark Shift $\chi\sim 0.1$~$\mu$eV $=V_\textrm{ex}/20$ (Appendix C).

\section{Two-Qubit Controlled-Phase Gate}
\begin{figure}
\includegraphics[width=8cm,height=5cm]{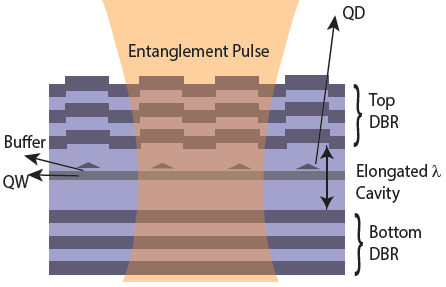}\hspace{-.5cm}\\
\caption{Illustration of the setup for a two-qubit operation. The pump pulse excites LPs in the trap regions directly below the QDs.}
\end{figure}
Figure 3 shows a schematic diagram for implementing two-qubit controlled-z operation between neighboring QD spins. A  $\sigma^+$ polarized laser pulse is incident normally on two adjacent QDs. The pulse is red-detuned by $\delta$ from the LP ground state at ${\bf{k}}_\|={\bf{0}}$. Furthermore, we assume a Gaussian profile for the pulse  $P(t)=P_0\exp(-t^2/\tau^2)$ applied between $-T_\mathrm{g}/2<t<T_\mathrm{g}/2$. If the tunnelling rate of polaritons between the traps is $U$, then the Hamiltonian for the LPs in the two neighboring traps is:
\begin{align}\begin{split}
H&=\delta\sum_{k=\ell,r}a^\dag_{1,k}a_{1,k}-\left(V_\mathrm{ex}+\chi\right)\sum_{k=\ell,r}a^\dag_{1,k}a_{1,k}\sigma_{\mathrm{z},k}\\
&+i\sqrt{\gamma_t}P(t)(a^\dag_{1,\ell}-a_{1,\ell}+a^\dag_{1,r}-a_{1,r})\\
&+U(a^\dag_{1,\ell}a_{1,r}+a^\dag_{1,r}a_{1,\ell})
\end{split}
\label{h2q}
\end{align}
where $a^\dag_{1,\ell(r)}, a_{1,\ell(r)}$ are the creation, annihilation operator for LPs with $J_\mathrm{z}=1$ in the left-$\ell$ (right-$r$) trap. For simplicity the two traps and QDs are assumed to be identical, but they need not be so. The detuning $\delta$ is large so that the LPs are only virtually excited in the trap, that is, the population of the LPs in the trap returns to zero at the end of the pulse at $t=T_\mathrm{g}/2$. By eliminating the LP modes, the effective Hamiltonian for the QD electron spins can be written as,
\be
H=-\frac{2|\alpha(t)|^2}{\delta+U}\left(V_\mathrm{ex}+\chi\right)^2\sigma_{\mathrm{z},\ell}\sigma_{\mathrm{z},r},
\label{gate}
\ee 
where,
\be
\alpha(t)=i\sqrt{\frac{\gamma_\mathrm{t}}{2}}\int_{-T_\mathrm{g}/2}^t P(t)\exp[-i(\delta+U)(T-s)]\mathrm{d}s.\nonumber\\
\ee
It is possible to choose the pump strength, duration and detuning, such that unitary corresponding to the Hamiltonian evolution in Eq.~\eqref{gate} is given by $U=(1,1,1,-1)$. Thus if $\gamma=0$ then it is possible to exactly implement a controlled phase gate between the two QD spins. For example, we find that with $\delta=5$ meV, $\chi=0.1$ $\mu$eV, $V_\mathrm{ex}=2$ $\mu$eV, $U=0.5$ meV, $\tau=9.3$~ns and peak pump power $\hbar\omega P_0^2=0.47$ W, it is possible to implement the gate in $T_\mathrm{g}=38$~ns. The number of LPs inside the resonators at the peak power is $\sim 400$. If $\gamma\neq 0$, then as seen by Eq.~\eqref{h2q} the fluctuations in the LP population leads to qubit dephasing at a rate given by,
\be
\gamma_\phi(t)\sim\gamma\frac{2|\alpha(t)|^2}{(\delta+U)^2}\left(V_\mathrm{ex}+\chi\right)^2
\ee
Assuming a cavity of quality factor $Q=76, 000$, we find that the photon decay rate $\gamma_\mathrm{p}=0.02$ meV. The relaxation rate of a QW exciton is $\gamma_\mathrm{x}\sim4$ $\mu$eV, so that the polariton decay rate $\gamma=|t^2_0|\gamma_\mathrm{p}+ |r|^2_0\gamma_\mathrm{x}=\frac{\gamma_\mathrm{p}}{2}+ \frac{\gamma_\mathrm{x}}{2}=0.012$ meV.  In this case, we find that the worst case fidelity of the C-phase gate to take an initial product state $\ket{\frac{1}{2},\frac{1}{2}}+\ket{\frac{1}{2},\frac{1}{2}}+\ket{-\frac{1}{2},\frac{1}{2}}+\ket{-\frac{1}{2},-\frac{1}{2}}$to the maximally entangled state $\ket{\frac{1}{2},\frac{1}{2}}+\ket{\frac{1}{2},-\frac{1}{2}}+\ket{-\frac{1}{2},\frac{1}{2}}-\ket{-\frac{1}{2},-\frac{1}{2}}$ is $99.4\%$.

\section{ Quantum Non-Demolition Measurement}
Next we will describe a scheme to achieve a fast, high fidelity, single-shot, QND measurement of the spin state of the QD electron. We discussed the importance of eliminating any back-action during the read-out process. The inherent spectral separation between the QD and QW excitations ensures that a probe pulse close to the QW LP resonance does not excite the QD single electron to trion or p-shell states. This eliminates the read-out backaction in the form of a spin-flip transition via excited states. We propose to use the QND readout mechanism introduced in reference 16. We will briefly describe the principle of the readout scheme. A horizontally (H) polarized probe laser, slightly detuned from the LP resonance ($\delta$), is incident normally over the QD whose electron spin is to be measured. 
It excites both $J_\mathrm{z}=1$ and $J_\mathrm{z}=-1$ LPs in the target trap. As described by Eq.~\eqref{hintred}, the degeneracy between the $J_\mathrm{z}=1$ and $J_\mathrm{z}=-1$ LPs is lifted due to Coulomb exchange interaction with the QD electron spin qubit. Consequently, the $J_\mathrm{z}=1$ and $J_\mathrm{z}=-1$ LPs evolve with different phases and amplitudes which is manifested by the introduction of a small vertically (V) polarized component in the light reflected from the cavity. The reflected light is elliptically polarized with its axis tilted by an angle $\propto \pm V_{\mathrm{ex}}$ (depending on whether $\ket{s_{\mathrm{z}}}=\ket{\pm \frac{1}{2}}$). The major difference between the setup in reference 16 and our scheme in the current work is that we now suggest the use of etched polariton traps to increase the exchange interaction between a QD electron and a polariton. Whereas in a planar cavity with no traps we estimated $V_\mathrm{ex}\sim 0.2$ $\mu$eV, with traps of radius $1$ $\mu$m, we estimate $V_\mathrm{ex}\sim 2$ $\mu$eV. With this improvement, we find that the measurement can now be performed an order of magnitude faster. Reference 16 also describes various possible sources of readout errors, namely: shot-noise error, phonon-assisted spin-flip scattering and radiative recombination of QD electron with QW holes. However, in the setup proposed here, there is an additional source of crosstalk error due to polariton tunnelling between neighboring traps, which is estimated in the supplement. For example, we find that in a single-sided cavity with $\gamma=0.027$ meV, it is possible to make a measurement in time $660$ ps with a total error of $\sim 0.1\%$ and peak number of photons $\sim 525$.


\section{ Single-Qubit Gates}
\begin{figure}
\includegraphics[width=6.5cm,height=3.5cm]{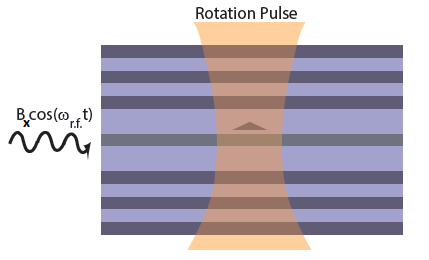}
\caption{The figure shows an in-plane r.f. magnetic field of small strength. In the absence of any optical pulse, the in-plane field is far-red detuned from the Zeeman splitting of the QD electron spin state. When an optical pulse is applied to excite LPs with $\langle a^\dag_{-1}a_{-1}\rangle=N$ such that $E_\mathrm{z}-V_\mathrm{ex} N$, the spin qubit rotates about the x-(or y-) axis. }
\label{onequbitcartoon}
\end{figure}
\begin{figure}
\includegraphics[width=6cm,height=5.5cm]{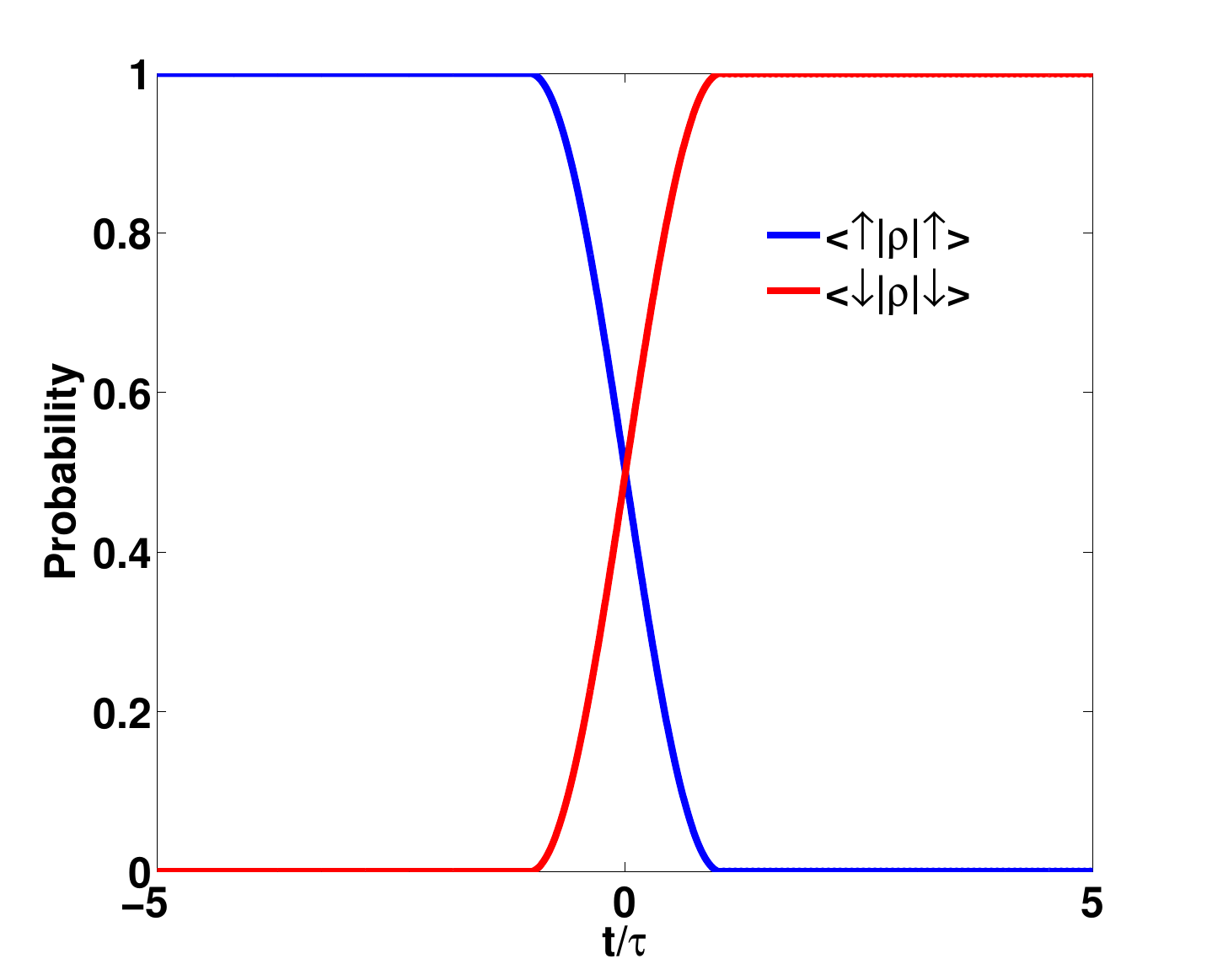}
\caption{Plot of probabilities $\langle\frac{1}{2}|\rho|\frac{1}{2}\rangle|^2$ and $\langle-\frac{1}{2}|\rho|-\frac{1}{2}\rangle|^2$ as a function of time during the application of a $\pi-$rotation pulse. }
\label{onequbit}
\end{figure}
Finally, in order to achieve universal quantum computation, we must be able to implement rotations of a single spin-qubit on the Bloch sphere. However, the quantum confinement effect restricts the polarization of the 2D-polaritons along the growth ($z$) axis \cite{winkler2008spin}. Thus, in order to achieve arbitrary qubit rotations we propose to use an external in-plane r.f. magnetic field (of magnitude $B_\mathrm{x}$ and frequency $\omega_\mathrm{r.f.}$) as shown in Fig.~\ref{onequbitcartoon}. The vertical magnetic field $B_0$ leads to a Zeeman splitting of $E_\mathrm{z}=g_e\mu_\mathrm{B} B_0$, where $g_e$ is the electron $g$-factor in the QD and $\mu_\mathrm{B}$ is the Bohr magneton.  The amplitude of the in-plane magnetic field $B_\mathrm{x}$ is such that $E_\mathrm{z}>\omega_\mathrm{r.f.}$ and $g_e\mu_\mathrm{B}B_\mathrm{x}\ll E_\mathrm{z}-\omega_\mathrm{r.f.}$. This ensures that, in the absence of an optical pump the in-plane magnetic field does not rotate the spin qubits. We propose to use an optical pump pulse to bring the qubit on resonance with the always-on r.f. magnetic field. The r.f. field performs the spin rotation \cite{gywat2004optical,kloeffel2013prospects} and the optical pulse performs the qubit addressing, which is an arrangement that has an analog in atomic physics with neutral-atom qubits \cite{wang2016single}. A similar approach has been used for single-qubit rotation of nuclear spins in silicon~\cite{kane1998silicon}, in which an externally applied voltage brings the nuclear spin in resonance with an in-plane a.c. magnetic field.  In contrast, here, a $\sigma^-$ pump pulse (red detuned from the LP resonance by $\delta$) is applied over the QD to excite $J_\mathrm{z}=-1$ LPs in the trap directly below it. The exchange interaction between the QD electron and $J_\mathrm{z}=-1$ LPs is represented by the Hamiltonian $H=-V_\mathrm{ex} a^\dag_{-1}a_{-1}{\sigma}_\mathrm{z}$. The key idea is to realize that this interaction leads to the modulation of the Zeeman energy of the QD electron spin by an amount $\propto -V_\mathrm{ex} a^\dag_{-1}a_{-1}$. Thus when $E_\mathrm{z}-V_\mathrm{ex} \langle a^\dag_1a_1\rangle=\omega_\mathrm{r.f.}$, the in-plane magnetic field rotates the spin along the $x$ axis. The angle by which the spin rotates is determined by length of the pump pulse and the rotation axis is determined by its phase. If the pump pulse is in-phase (out-of-phase) with the in-plane magnetic field, then the qubit rotates about the $x(y)$ axis. To illustrate, consider a rotation pulse $F_\mathrm{\tiny{T}}$, red-detuned from the LP by $\delta=6$ meV with pulse shape,
\begin{align}\begin{split}
F_\mathrm{\tiny{T}}&=F_0 e^{-\frac{|t+\tau|^2}{\tau_\mathrm{\small{r}}^2}} \quad \forall \quad t<-\tau,\\
&=F_0 \quad \quad \quad \forall  -\tau\leq 0\leq\tau,\\
&=F_0 e^{-\frac{|t-\tau|^2}{\tau_\mathrm{\small{r}}^2}} \quad \forall \quad t>\tau,
\end{split}
\label{pulse1}
\end{align}
where $F_0=769.8$ $(\sqrt{\mathrm{meV}})^{-1}$, $\tau=200$ ps and $\tau_\mathrm{\small{r}}=5$ ps. The value of $F_0$ is chosen so that in a symmetric two-sided cavity with $\gamma=0.027$ meV,  the average number of LPs at peak power is $\langle a^\dag_{-1}a_{-1}\rangle=N=500$ and thus, $V_\mathrm{ex}N=1$ meV (for $V_\mathrm{ex}=2$ $\mu$eV). The in-plane magnetic field is of magnitude $B_\mathrm{x}=0.023$ mT and frequency such that $E_\mathrm{z}-\omega_\mathrm{r.f.}=1$ meV. This results in the $\pi$ rotation of the spin-qubit in $\sim 420$ ps. Figure~\ref{onequbit} plots the probability for the spin to be in one of the two states, $\langle\frac{1}{2}|\rho(t)|\frac{1}{2}\rangle|^2$ and $\langle-\frac{1}{2}|\rho(t)|-\frac{1}{2}\rangle|^2$ during the application of a $\pi$ rotation pulse. Like the two-qubit gate the fidelity of this gate is limited by the dephasing caused by the photon number fluctuations. These fluctuations induce uncertainty in the energy of the qubit, leading to decoherence and we estimate the single-qubit gate error probability in our scheme to be $=0.2\%$ (Appendix F).

\section{Conclusions}
To summarize, Table I lists the theoretical gate times and average fidelities that can be achieved in QD spin qubits using the proposed technique of indirect optical control mediated by QW LPs in lithographically-defined traps. The table also shows the performance of the best non-polariton-based optical schemes. The polariton-based two-qubit gate scheme requires a cooperativity $C$ that is substantially lower than that needed by the purely optical schemes (by a factor of $\sim 100$ for $F=99.8\%$). Furthermore we found that the use of trapped polaritons increases the exchange interaction between QD electrons and QW polaritons by a factor of 10 (from $\sim 0.2 ~\mu \textrm{eV}$, as calculated in Ref. ~\cite{puri2014single}, to $\sim 2 ~\mu \textrm{eV}$) under otherwise identical conditions, and that the two-qubit gate is consequently $100\times$ faster. We have also shown how it may be possible to achieve a single-qubit rotation in Faraday geometry within $\sim 420$ ps and with a fidelity of $99.9\%$. An improvement in the QND measurement time of a factor of $30$ is achieved, but more importantly, the probability of error is reduced by a factor of $\sim 200$, versus the non-trap polariton-based scheme analyzed in Ref. ~\cite{puri2014single}. The principle differences between the scheme described here and the non-trap polariton scheme have also been summarized in table II. In conclusion, we have presented a substantial modification to previous polariton-mediated two-qubit gate schemes, namely the introduction of traps, and have shown how high-fidelity and fast two-qubit and one-qubit gates, as well as single-shot QND measurement, can be performed in this setting. Our results suggest that polariton-based approaches to controlling quantum-dot spin qubits may ultimately be able to reach the scalability and fidelity requirements for constructing fault-tolerant quantum processors based on the surface code.
\begin{widetext}
                                                                                                                                                                                                                                                        
\begin{table}[h]
\begin{widetext}
\caption{Comparison of the theoretical gates times and fidelities that can be achieved with QD electron qubits using the proposed technique of indirect optical control mediated by QW LPs with the existing optical schemes.}
\begin{tabular}{||c|c|c|c|c||}
\hline
\hline
& \multicolumn{2}{c} {Non-QW-LP Optical Schemes} &\multicolumn{2}{|c||}{Proposed Scheme with QW-LPs}\\\hline
& Gate Time&Fidelity & Gate Time &Average Gate Fidelity\\\hline
Two-qubit gate & 100 ns~\cite{ladd2011simple} &$99.9\%$~\cite{ladd2011simple}& 24 ns & 99.8$\%$\\ \hline
$C$ & \multicolumn{2}{c} {$10^3$} &\multicolumn{2}{|c||}{$20$}\\\hline
Single-qubit operation & \multicolumn{2}{c|}{Not possible in Faraday geometry,} & 420  ps & 99.7$\%$\\
& \multicolumn{2}{c|}{in QDs using optical pulses and no local wiring }& & \\
& \multicolumn{2}{c|}{(but possible in gate defined QDs).}& & \\\hline
QND measurement & 20 ns~\cite{gazzano2013bright,delteil2014observation} & 95$\%$~\cite{gazzano2013bright,delteil2014observation}& 660 ps & 99.97$\%$\\ \cline{4-5}
  &  & &\multicolumn{2}{c||}{Single-sided cavity}\\\hline\hline
\end{tabular}
\end{widetext}
\end{table} 
\end{widetext}

\begin{widetext}

\begin{table}[h]
\begin{widetext}
\caption{Principle differences between the proposed technique of QW LPs confined in traps and previous approaches without the trap.}
\begin{tabular}{||c|c|c||}
\hline
\hline
 & Previous QW-LP approach & QW-LP approach with traps\\ 
 \hline
 Exchange interaction $V_\textrm{ex}$ & 0.2 $\mu$eV & 2 $\mu$eV\\ \hline
 Universal gate set & Not possible & Possible\\ 
 & (Two-qubit gate and QND measurement & (All operations in Faraday geometry)\\
 & in Faraday geometry, while single-qubit & \\ \hline
 & rotation in Voigt geometry) &\\ \hline
 Other effects & Direct optical electron-trion transition & Two-qubit gate control pulse takes \\
 &not considered  &  into account direct optical\\ 
 &  & electron-trion transition\\ \cline{3-3}
 & & \multicolumn{1}{c||}{Estimate of crosstalk due to}\\
 & & \multicolumn{1}{c||}{LP tunnelling between traps}
 
 \\\hline\hline
\end{tabular}
\end{widetext}
\end{table} 
\end{widetext}

\section{Acknowledgements}
This work was supported by the JSPS through its FIRST Program. We thank Kai M\"uller (TU M\"unchen) for suggesting several references on one-qubit control using direct radiofrequency manipulation.

\bibliography{PuriPRB_v3.bbl}{}
 
\newpage

\appendix

\section{Polariton Tunnel Coupling Between Neighboring Traps}

\begin{figure}
\includegraphics[width=4.5cm,height=4cm]{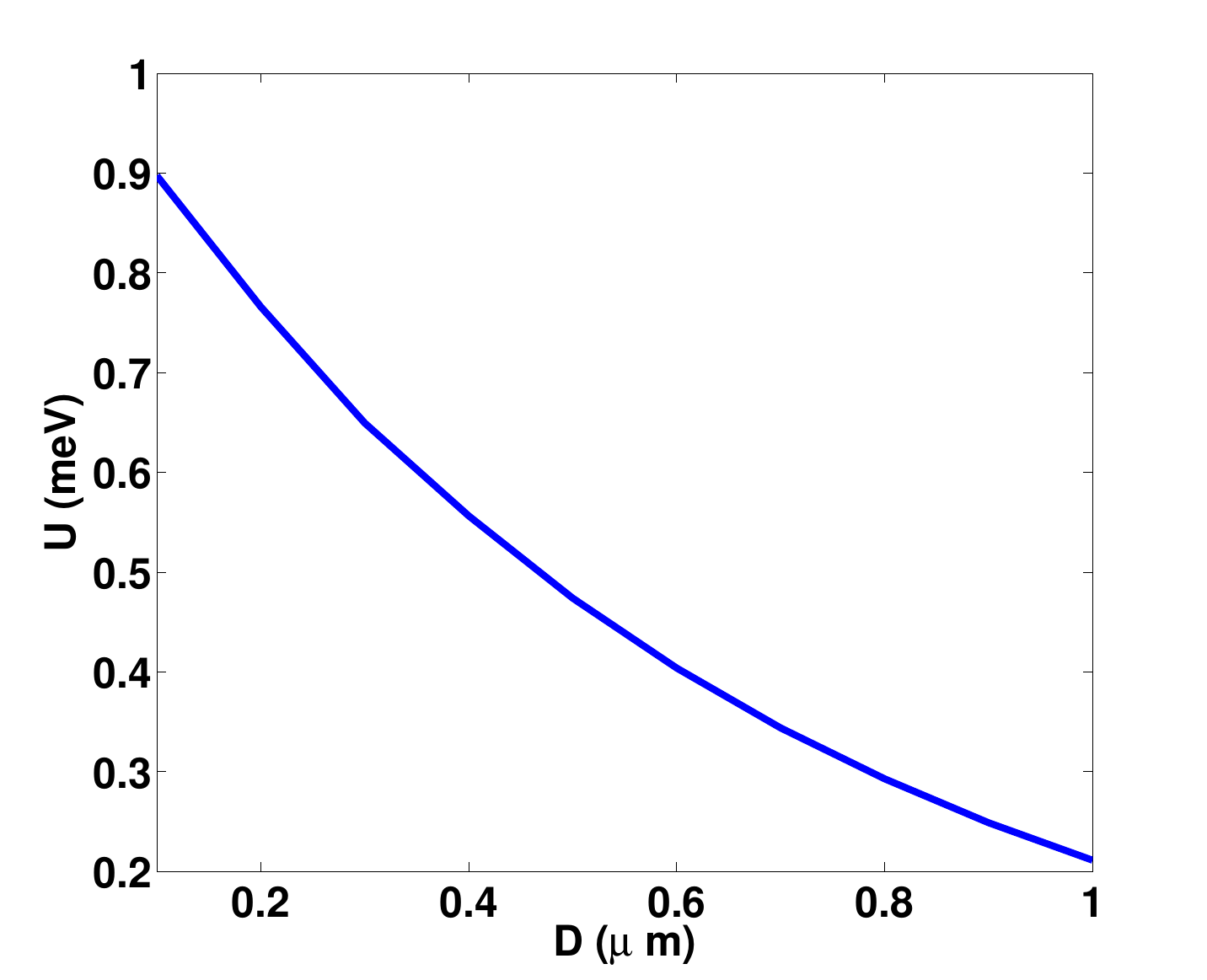}\includegraphics[width=4.5cm,height=4cm]{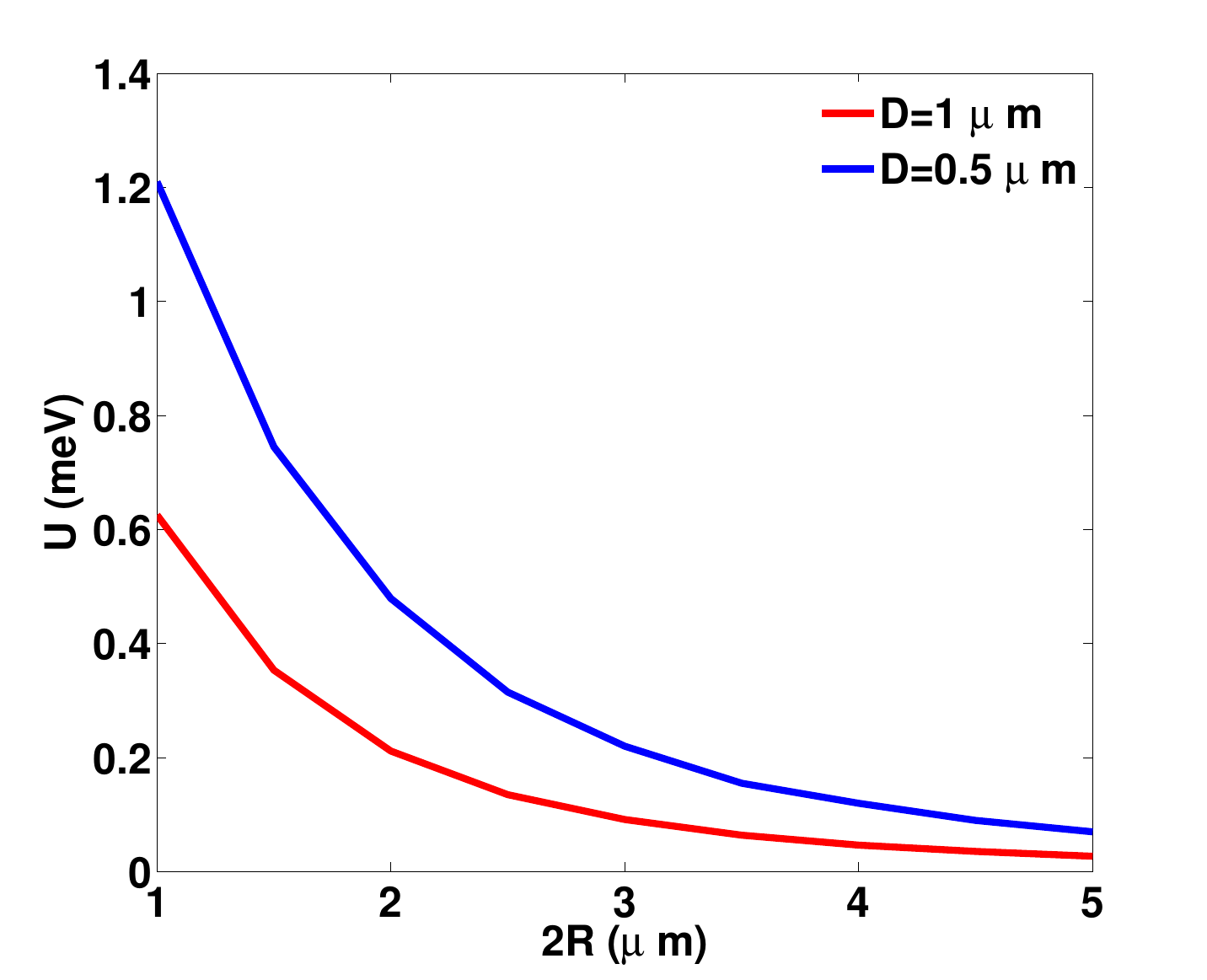}\\
\hspace*{.2cm} (a) \hspace{4cm} (b)
\caption{Variation of $U$ with (a) $D$ for $2R=3$ $\mu$m (b) $2R$ for $D=1$ $\mu$m (red curve) and $D=0.5$ $\mu$m (blue curve).\label{variation}}
\end{figure}

\begin{figure}
\includegraphics[width=4.5cm,height=4cm]{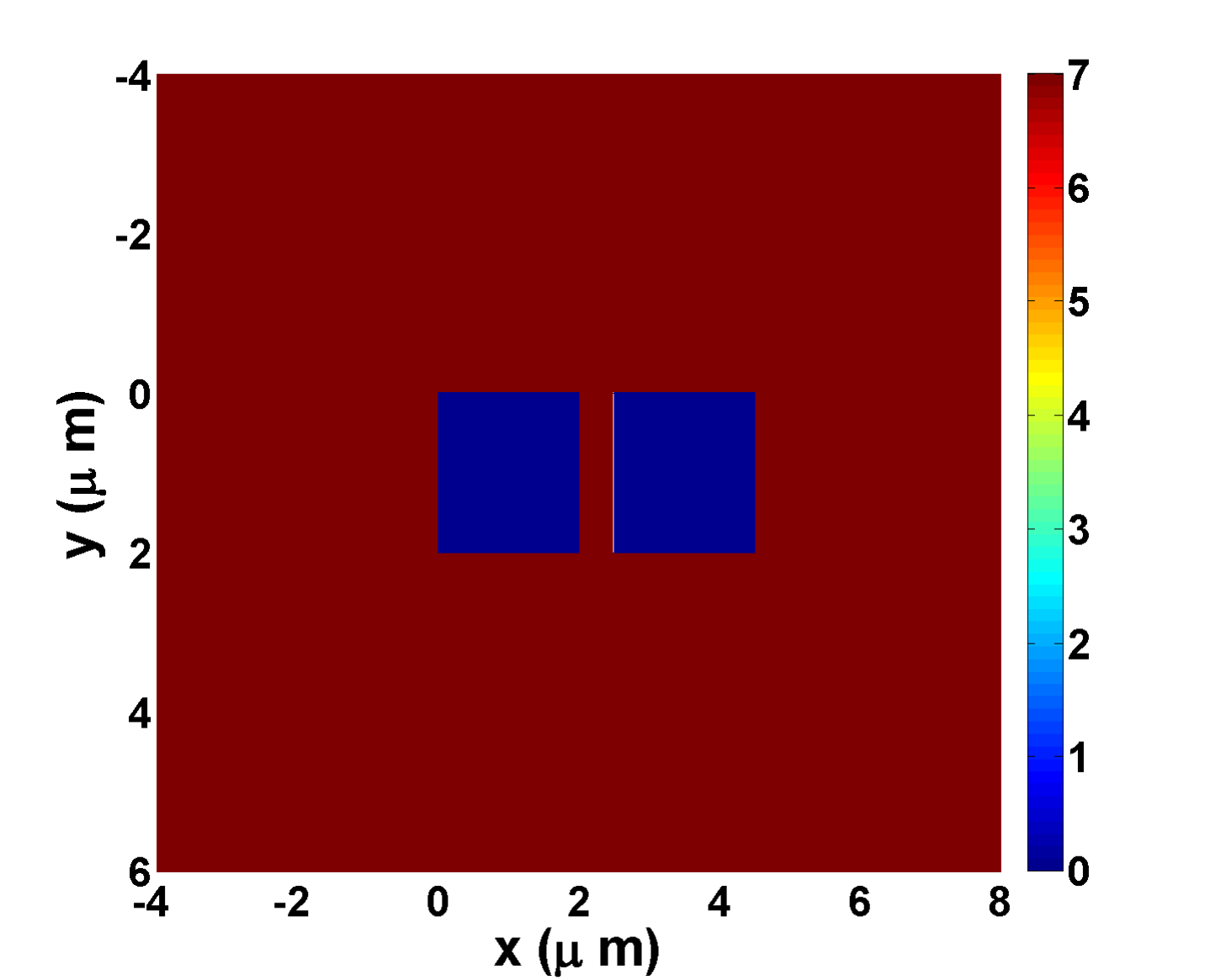}\\
\hspace{0cm} (a)\\
\includegraphics[width=4.5cm,height=4cm]{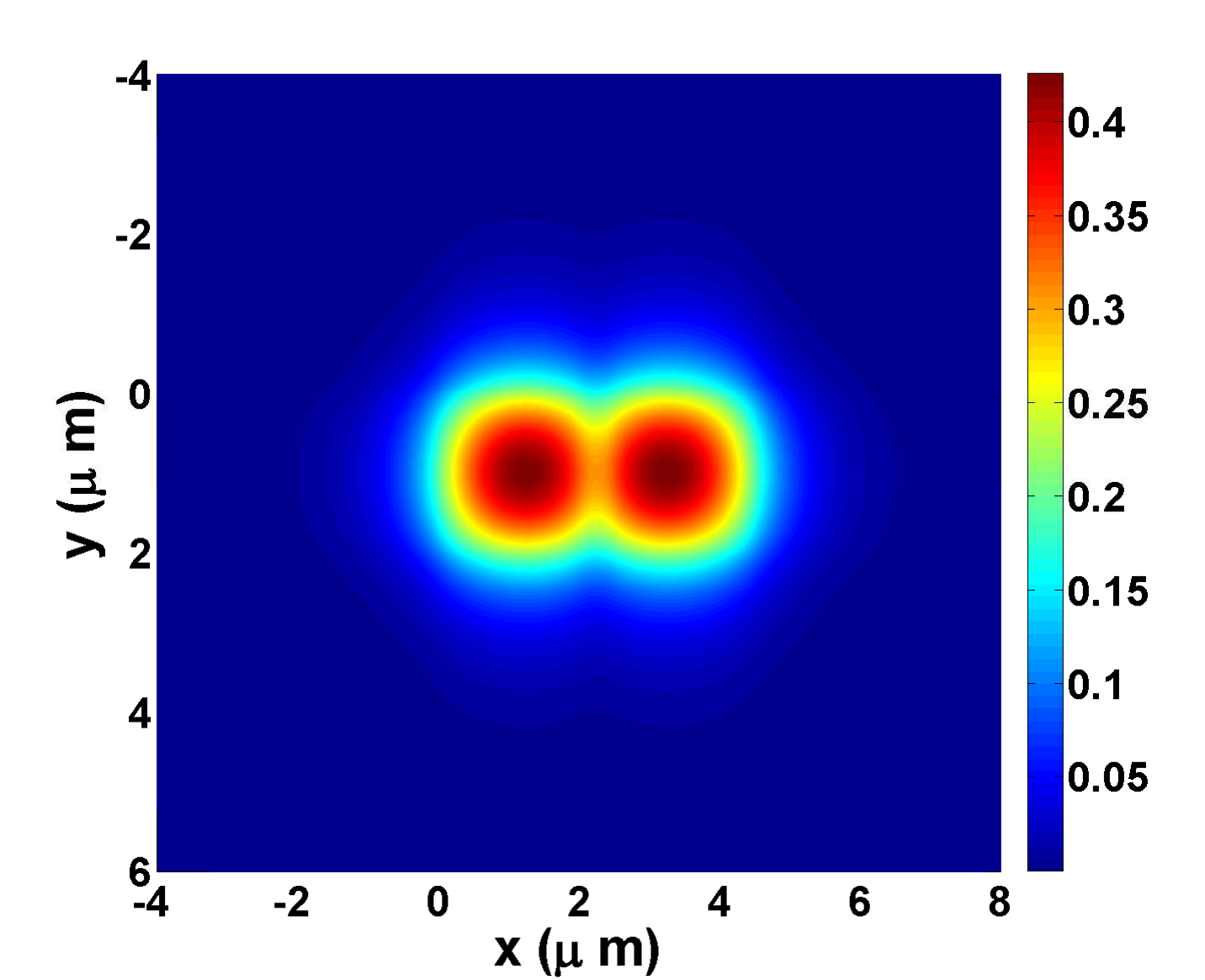}\includegraphics[width=4.5cm,height=4cm]{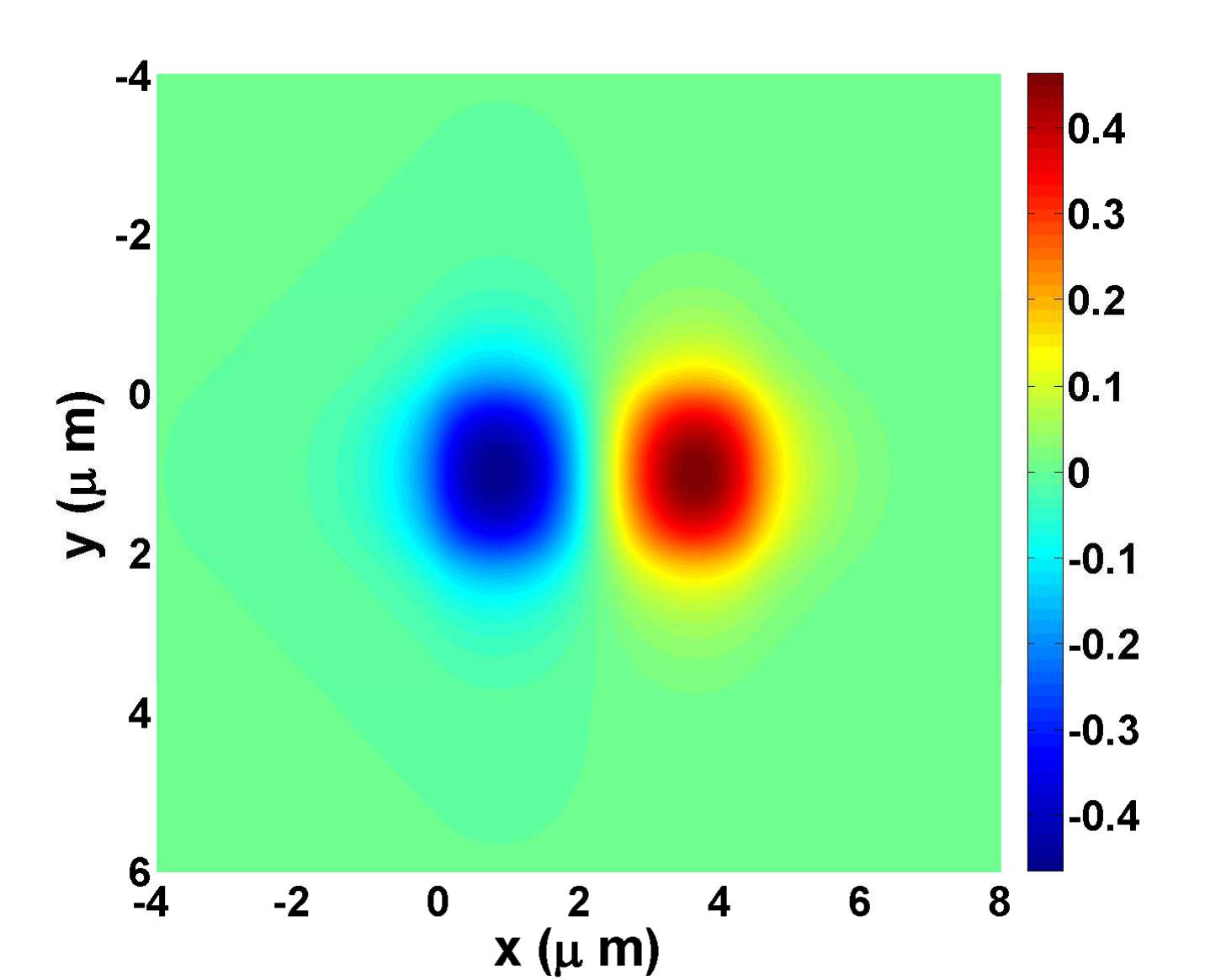}\\
\hspace{0cm} (b)\hspace{4cm} (c)
\caption{(a) Illustration of the coupled LP traps, modeled as two square potential wells of depth 7 meV with sides of length $2R$ $(=2$ $\mu$m) and separated by $D$ (=0.5 $\mu$m). (b) and (c) Normalized ground state and excited state wavefunction of the LP in the coupled traps.  \label{wavefunc}}
\end{figure}

In this section we outline the method used to estimate the tunnel coupling constant ($U$) between neighboring traps. Our method is motivated from the analysis in reference~\cite{sarchi2008coherent}. The main paper discussed how the local modulation of the cavity length created traps for polaritons. In order to estimate $U$, we model the traps as square potentials of depth $\sim 7$ meV and sides of length $2R$. If the normal mode splitting ($2\hbar\Omega_\mathrm{R}$) between the upper polaritons (UPs) and lower polaritons (LPs) is larger than the coupling constant $U$, then the LPs at ${\bf{k}}_\|={\bf{0}}$ can be treated as quasi-particles of mass $m_\mathrm{LP}$. Typically, $2\hbar\Omega_\mathrm{R}\sim 3-4$ meV and when the cavity photons are resonant with the quantum well (QW) excitons at ${\bf{k}}_\|={\bf{0}}$, then $m_\mathrm{LP}\sim 4\times 10^{-5}m_0$ ($m_0$ is the mass of free electron). We can numerically evaluate the ground-state ($E_\mathrm{gs}$) and first excited-state ($E_\mathrm{es}$) energies of a LP (of mass $m_\mathrm{LP}$) in coupled square potential wells (each of depth 7 meV) by solving the time-independent Schr\"{o}dinger equation:
\be
\left(-\frac{\hbar^2\nabla^2}{2m_\mathrm{LP}}+U(x,y)\right )\Psi(x,y)&=&0,
\ee
with the coupled square potential illustrated in Fig. 7(a) and given by:
\be
U(x,y)&=&0, \quad 0<x<2R \quad \& \quad 0<y<2R\nonumber\\
&=&0, \quad 2R+D<x<4R+D \quad \& \quad 0<y<2R\nonumber\\
&=&7 \quad \mathrm{meV, everywhere else}
\ee
Finally, the linear tunnel coupling strength is estimated as:
\be 
U=\frac{1}{2}(E_\mathrm{es}-E_\mathrm{gs})
\label{cops}
\ee 
Figure~\ref{variation}(a,b) shows the dependence of $U$ on $D$ and $R$ and the ground state and excited state wavefunctions for D=0.5 $\mu$m and $R=1$ $\mu$m are illustrated in Fig.~\ref{wavefunc}(a,b). For these parameters and using Eq.~\eqref{cops}, we estimate $U\sim 0.5$ meV.

\section{Selective excitation of LP in only one trap}
\begin{figure}
\includegraphics[width=4.5cm,height=4cm]{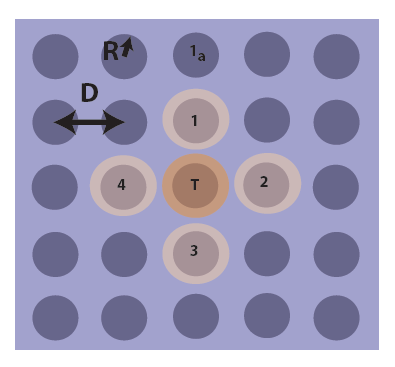}
\caption{Top view of the structure showing the LP traps (blue circles) and pump laser pulses over the target trap, T (($F_\mathrm{\tiny{T}}$) and nearest-neighbor traps 1-4 ($F$). As shown in the main text when $F=\frac{iF_\mathrm{\tiny{T}} U}{i\delta+\frac{\gamma}{2}}$, LPs are only excited in the target trap.}
\end{figure}
For clarity in presentation, we will only consider the nearest-neighbor coupling. Consider a target trap T, which is directly under the QD that hosts the spin qubit we intend to manipulate. The neighboring traps are numbered $1-4$ and our aim is to excite a ${\bf{k}}_\|={\bf{0}}$ LP mode only in trap T (Fig. 2(c)). In order to do so we excite LPs in trap T and the four neighboring traps  by a pump $F_\mathrm{\tiny{T}}$ and $F$ respectively. Both the pumps are red detuned from the ${\bf{k}}_\|={\bf{0}}$ LP mode in the traps by $\delta$. In the rotating frame of the pump pulse, the rate equations for the coherent amplitude ($\alpha_\mathrm{\tiny{T(1-4)}}$) of the ${\bf{k}}_\|={\bf{0}}$ LP mode in trap T(1-4) are given by:
\be
\frac{d\alpha_\mathrm{\tiny{T}}}{dt}&=&\sqrt{\gamma_t}F_\mathrm{\tiny{T}}-\left(i\delta+\frac{\gamma}{2}\right)\alpha_\mathrm{\tiny{T}}-iU\sum_{i=1}^4\alpha_i,\\
\label{trapt}
\frac{d\alpha_i}{dt}&=&\sqrt{\gamma_t}F-\left(i\delta+\frac{\gamma}{2}\right)\alpha_i-iU\alpha_\mathrm{\tiny{T}} ,
\label{trapn}
\ee
where, $i=1,2,3,4$, $\gamma (\gamma_t)$ is the decay rate of polaritons from the cavity (top mirror), $|F_\mathrm{\tiny{T}}|^2$ ($|F|^2$) is the photon flux in the pump incident on trap T (neighboring traps). It will be instructive to solve the above equations at steady state, under the assumption that $|\alpha_\mathrm{\tiny{T}}|\gg |\alpha_i|$. Under these conditions, the steady state amplitude of the LP mode in trap T is:
\be
\alpha_\mathrm{\tiny{T}}=\frac{\sqrt{\gamma_t}F_\mathrm{\tiny{T}}}{i\delta+\frac{\gamma}{2}}.
\label{a}
\ee
By substituting Eq.~\eqref{a} into Eq.~\eqref{trapn} we see that if $F={i F_\mathrm{\tiny{T}} U}/({i\delta+{\gamma}/{2}})$, then the effective pump term for LPs in the neighboring traps vanishes. 
\section{Estimation of Stark Shift}
The dipole coupling content is given by~\cite{andreani1999strong},
\be
g=\sqrt{\frac{e^2 f}{4\epsilon m_0 V}}
\ee
where, $f$ is the oscillator strength. For $f \sim 50$~\cite{reithmaier2004strong} and mode volume $V=0.5$ $\mu$m$^3$ (for trap radius $R=1$ $\mu$m), the coupling constant $g\sim 60$ $\mu$eV for an. Thus, if a pulse is applied close to the LP resonance and $t_0=\frac{1}{\sqrt{2}}$, then the Stark shift will be $\chi= {(60 \mu\mathrm{eV})^2}/{2 \times 20~\textrm{meV}}=0.1$~$\mu$eV $=V_\textrm{ex}/20$.

\section{Cavity Mode Volume}
The mode volume of the cavity is calculated by first decoupling the field along the growth ($z-$axis) and in the cavity plane. This approximation is valid because of strong confinement of the photon modes along the growth direction. The mode volume then becomes $V=AL$, where $A$ is the mode area and $L$ is the mode length of the vertical cavity. The in-plane mode in a single trap can be calculated following the procedure in section I of the supplement and is approximated by the Gaussian $\psi_\|=A_0e^{-\frac{r^2}{a^2}}$, where $a=1.2$ $\mu$m. The effective cavity length $L_\mathrm{c}$ is a few wavelengths longer than the cavity length $\lambda$, due to the penetration of the cavity field into the DBR:
\be
L_\mathrm{c}&=&\lambda+L_\mathrm{DBR},\nonumber\\
L_\mathrm{DBR}&=&\frac{\lambda}{2} \frac{n_1n_2}{n_\mathrm{c}|n_1-n_2|},\nonumber 
\ee
where $n_1,n_2(n_\mathrm{c})$ are the refractive index for the materials of the DBR (cavity). For a GaAs/AlGaAs DBR mirrors and GaAs cavity, $n_1=3$, $n_2=n_\mathrm{c}=3.6$. The mode profile along the growth direction will be $\psi_\bot=B_0\sin\left(\frac{\pi z}{L_\mathrm{c}}\right )$. Thus, the mode area and mode length are:
\be
A=\frac{\int^\infty_0 |\psi_\||^2 d^2{\bf{r}}}{|\psi_\||^2_\mathrm{max}}, &\quad& L=\frac{\int^\infty_0 |\psi_\bot|^2 dz}{|\psi_\bot|^2_\mathrm{max}}\nonumber\\
\Rightarrow A=\frac{\pi a^2}{2}, &\quad& L=\frac{L_\mathrm{c}}{2}
\ee
Thus, the mode volume in the $\lambda$(=910 nm) cavity is $V=0.5$ $\mu$m$^3$.

\section{Crosstalk due to tunnel coupling with neighboring traps}
In section III we outlined a scheme to minimize the coupling of LPs in neighboring traps. Despite this, some LPs are injected in the neighboring traps. These LP modes interact with the neighboring electron spin qubits, leading to decoherence. Consider the setup described in section III (Fig. 2(c)). Suppose, the  spin state of the electrons trapped in the QDs over the target trap is $\ket{\frac{1}{2}}_\mathrm{\tiny{T}}$ and spin state in the neighboring traps is the maximally entangled state $\ket{\psi}=\frac{\ket{\frac{1}{2}_1,\frac{1}{2}_2,\frac{1}{2}_3....}+\ket{-\frac{1}{2}_1,-\frac{1}{2}_2,-\frac{1}{2}_3....}}{\sqrt{2}}$. We choose this state for the neighboring qubits as it will experience maximum decoherence. After the application of the measurement pulses $F$ and $F_\mathrm{\tiny{T}}$, the spin state of the target qubit will be projected on to the state $\ket{\frac{1}{2}}_\mathrm{\tiny{T}}$. But the state of the neighboring qubits will be a mixed state represented by the density matrix $\rho_\mathrm{f}$. The error introduced in the states of neighboring qubits during a measurement of the target qubit is: $P^\mathrm{c}_\mathrm{e}=1-F$, where $F=\mathrm{Tr}[\sqrt{\rho_\mathrm{f}}\ket{\psi}\bra{\psi}\sqrt{\rho_\mathrm{f}}]$ is the fidelity. In the example presented in table II, if a measurement is made in a single-sided (symmetric two-sided) cavity for $\tau_\mathrm{meas}=660(750)$ ps with $F_\mathrm{\tiny{T}}=41.1(70.8)$ $1/\sqrt{\mathrm{ps}}$ at $\delta=0.3$ meV, then the probability of error introduced in the neighboring spin state is $P^\mathrm{c}_\mathrm{e}\sim 0.002 (0.004)\%$.

\section{Effective Hamiltonian for the two-qubit gate operation}
Consider the Hamiltonian for the two-qubit gate operation in Eq.~\eqref{h2q},
\begin{align}\begin{split}
H&=\delta\sum_{k=\ell,r}a^\dag_{1,k}a_{1,k}-\sum_{k=\ell,r}\left(V_{\mathrm{ex},k}+\chi_k\right)a^\dag_{1,k}a_{1,k}\sigma_{\mathrm{z},k}\\
&+i\sqrt{\gamma_t}P(t)(a^\dag_{1,\ell}-a_{1,\ell}+a^\dag_{1,r}-a_{1,r})\\
&+U(a^\dag_{1,\ell}a_{1,r}+a^\dag_{1,r}a_{1,\ell})
\end{split}
\end{align}
where, $\left(V_{\mathrm{ex},k}+\chi_k\right)$ is the coupling strength between the $k=\ell,r$ QD electron spin with the LPs in the traps below the respective QDs. $U$ is the tunnel coupling between the traps and $P(t)$ is the rate at which the LPs are injected in the trap. Making the transformation $a_{1,\ell}=(a+b)/\sqrt{2}$ and $a_{1,r}=(a-b)/\sqrt{2}$, the above Hamiltonian reduces to,
\begin{align}
\begin{split}
H'&=(\delta+U)a^\dag a+(\delta+U)b^\dag b+i\sqrt{\frac{{\gamma_t}}{{2}}}P(t)(a^\dag+a)\\
&-\left(V_{\mathrm{ex},\ell}+\chi_\ell\right)(a^\dag a+b^\dag b+a^\dag b+b^\dag a)\sigma_{\mathrm{z},\ell}\\
&-\left(V_{\mathrm{ex},r}+\chi_r\right)(a^\dag a+b^\dag b-a^\dag b-b^\dag a)\sigma_{\mathrm{z},r}
\end{split}
\end{align}
Note that in this form, the external field $P(t)$ only drives the mode $a$ while the mode $b$ remains undriven. As a result, $\langle b\rangle=\langle b^\dag b\rangle=0$ and by eliminating the mode $b$, the above Hamiltonian reduces to,
\begin{align}
\begin{split}
H'&=(\delta+U)a^\dag a+i\sqrt{\frac{{\gamma_t}}{{2}}}P(t)(a^\dag+a)\\
&-a^\dag a[\left(V_{\mathrm{ex},\ell}+\chi_\ell\right)\sigma_{\mathrm{z},\ell}+\left(V_{\mathrm{ex},r}+\chi_r\right)\sigma_{\mathrm{z},r}]
\end{split}
\end{align}
On this Hamiltonian, we now apply a spin-dependent displacement transformation $D(\alpha')=\exp(i\alpha' a^\dag-i\alpha'^* a)$ where $\alpha'=\alpha-[(V_{\mathrm{ex},\ell}+\chi_\ell)\sigma_{\mathrm{z},\ell}+(V_{\mathrm{ex},r}+\chi_r)\sigma_{\mathrm{z},r}]\alpha/(\delta+U)$ so that effective Hamiltonian becomes,
$H_\mathrm{eff}=D^\dag(\alpha')H'D(\alpha')-iD^\dag(\alpha')\dot{D}(\alpha')$. If we now choose $\dot{\alpha}=-i(\delta+U)\alpha+i[(V_{\mathrm{ex},\ell}+\chi_\ell)\sigma_{\mathrm{z},\ell}+(V_{\mathrm{ex},r}+\chi_r)\sigma_{\mathrm{z},r}]\alpha+\sqrt{\gamma_t/2} P(t)$, then the effective Hamiltonian for the qubits reduces to,
\begin{align}
H_\mathrm{eff}&=-\frac{2|\alpha(t)|^2}{\delta+U}\left(V_{\mathrm{ex},\ell}+\chi_\ell\right)\left(V_{\mathrm{ex},r}+\chi_r\right)\sigma_{\mathrm{z},\ell}\sigma_{\mathrm{z},r},
\end{align}
where,
\be
\alpha(t)\sim i\sqrt{\frac{\gamma_\mathrm{t}}{2}}\int_{-T_\mathrm{g}/2}^t P(t)\exp[-i(\delta+U)(T-s)]\mathrm{d}s.\nonumber\\
\ee
and we have simplified the expression for $\alpha$ to zeroth order in the qubit-polariton interaction energy. Without loss of generality it is then possible to choose $\left(V_{\mathrm{ex},\ell}+\chi_\ell\right)=\left(V_{\mathrm{ex},r}+\chi_r\right)$ to recover the Eq.~\eqref{gate}

\section{Dephasing error in the two- and single-qubit gate schemes}
The two and single qubit gate schemes rely on the electron spin qubit-polariton interaction which leads to a polariton-number dependent shift in the energy of the qubit. As a result the quantum fluctuations in the polariton number population or in other words shot-noise, leads to dephasing of the qubit~\cite{gambetta2006qubit}. In a single-qubit gate the interaction Hamiltonian is given by $-V_\mathrm{ex}a^\dag_{-1}a_{-1}\sigma_\mathrm{z}$, which leads to dephasing at rate $\gamma_\phi(t)=2V_\mathrm{ex}^2\langle\delta n(t)\delta n(t')\rangle dt'=2V_\mathrm{ex}^2\gamma|\alpha(t)|^2/\delta^2$ where $n=a^\dag_{-1} a_{-1}$ and $\delta n(t)=n(t)-\langle n(t)\rangle$. In a two-qubit gate the interaction Hamiltonian between the mode $a$ and the qubits is given by $-\left(V_\mathrm{ex}+\chi\right)a^\dag a(\sigma_{\mathrm{z},\ell}+\sigma_{\mathrm{z},r})$ and the joint dephasing rate of the qubits because of photon shot noise of mode $a$ is $2\left(V_\mathrm{ex}+\chi\right)^2\gamma|\alpha(t)|^2/(\delta+U)^2$.

\end{document}